\documentclass[aps,showpacs,
twocolumn
]{revtex4} %\documentstyle[12pt]{ioplppt}
\usepackage[cp1251]{inputenc}
\usepackage{graphicx}
\begin{document}

\title{\large \bf
HTSC-glue in doped copper oxides and iron pnictides:\\ mobile
CT-excitons within in-plane Ginzburg HTSC-sandwich }
\author{L.S.Mazov}
\affiliation{Institute for Physics of Microstructures, Russian
Academy of Sciences, Nizhny Novgorod, 603600 Russia}
\begin{abstract}
It is demonstrated that high critical temperature of
superconducting transition in cuprates and new iron-based
superconductors is reached because of the Little-Ginzburg exciton
mechanism of HTSC when Cooper pairing of mobile charge carriers is
mediated by excitons which characteristic energy is essentially
higher than Debye one for phonons. The effectiveness of such
mechanism in these doped compounds is provided due to a series of
planar Ginzburg HTSC-'sandwiches': 'insulator'-'metal'-
'insulator' (stripe structure) naturally forming in conducting
planes below the onset pseudogap temperature in the normal state.
The parameters of mobile, planar charge-transfer (CT) excitons in
outer 'insulating' plates of such in-plane HTSC-'sandwich' are
exactly within the optimal range predicted by Ginzburg about forty
years ago.
\end{abstract}
 \maketitle {\bf 1.
Introduction.} The discovery of superconductivity in iron-based
compounds \cite{Hos08} permits one to consider them as a new
(copper-free) class of high-$T_c$ superconductors. Indeed, there
are some difference of detailed properties of concrete compounds
due to features of their band structure, but they exhibit a number
of another properties which bring them together with cuprates, for
example, layered structure, transition to a class of
superconductors under doping of parent, nonsuperconducting
compounds etc. (for review, see \cite{Hos09,Ginz09}). However, in
spite of a large number of works appearing since discovery of new
iron-based superconductors ($\sim 100$ papers per month that is of
course less than, but compared with, speed of appearance of
publications on the study of properties of cuprates: $\sim 400$
papers per month, in first ten years after their discovery
(see,e.g., \cite{Ginz04}), mechanism of superconductivity in new
iron-based superconductors (as well as in cuprates) remains to be
unclear (cf. \cite{Ginz09,Hos09}). In this work, there are
presented the evidences indicating that reaching of high
temperature for superconducting transition in cuprates in 1986/87
(and in iron-based compounds in 2008) appears to be demonstration
of successful realization of the old idea by W.A.Little
\cite{Little64} and V.L.Ginzburg \cite{Ginz64} on synthesis of
high-temperature superconductors with exciton mechanism of
attraction between conduction electrons. Moreover, the picture
obtained evidents also that this mechanism is first realized
effectively, just as it was predicted by Ginzburg (see, e.g.
\cite{Ginz6870,Ginz76,PHTSC}), in inhomogeneous system consisting
of three layers (HTSC-sandwich): insulator-metal-insulator (which
is presented in these compounds in planar form), when conduction
electrons (mobile charge carriers) in metallic spacer of sandwich
are paired near the Fermi surface due to interaction with excitons
in outer insulating plates at temperature which exceeds
significantly the phonon limit ($T_c^{ph} \le 40$ K, for details,
see, e.g., \cite{Ginz6870,Ginz76,PHTSC}).\\[1mm] {\bf 2. Exciton
mechanism of high-$T_c$ superconductivity: the sandwich-like
systems.} As known, in exciton mechanism of HTSC, proposed in
pioneer works by Little \cite{Little64} and Ginzburg \cite{Ginz64}
to increase critical temperature of superconducting transition
$T_c$, the exhange of conduction electrons by virtual phonons at
Cooper pairing is replaced to exchange by virtual excitons which
characteristic energy $\hbar \omega_{ex}$ is essentially higher
than Debye one $\hbar \omega_{D}$, characteristic for phonon
mechanism. The consideration is performed in frames of BCS model
in which the critical temperature is determined as (see, e.g.
\cite{Ginz6870}) $$ T_{c} = \Theta \,exp\,(-{1}/{g}). \eqno(1) $$
where $\Theta \approx \Theta_{ex} = \hbar \omega_{ex}/k_B$ is the
characteristic temperature of the excitons, $\omega_{ex}$ is the
characteristic frequency of excitons, $g = N(\varepsilon_F)V$ is
the effective constant of interaction, $N({\varepsilon}_F)$ is the
density of electronic states at the Fermi level $\varepsilon_F$ in
the normal state, $V$ is the matrix element of interaction. As it
was demonstrated in \cite{Ginz6870,Ginz76,PHTSC}, the following
set of parameters is optimal to obtain high-temperature
superconductivity on the basis of the exciton mechanism $$ \left\{
\begin{array} {l} \Theta_{ex} = {\hbar \omega_{ex}}/{k_B} \sim
10^{3} \div 10^{4} K
\\[2mm]  g = g_{ex} \ge {1}/{5} \div {1}/{3}
\end{array}\right.
\eqno(2) $$

Further, in works by Ginzburg (see, e.g. \cite{Ginz6870,PHTSC}),
it was concluded that because of strong excitonic damping in usual
three dimensional (3D) metals, the using of inhomogeneous systems:
one dimensional (1D) (see \cite{Little64}) or two dimensional (2D)
\cite{Ginz64,Ginz6870,PHTSC}, is most effective for realization of
exciton mechanism. The preference of 2D-systems is, in particular,
in smaller level of fluctuations as compared with 1D case. From
results of analysis (see monography \cite{PHTSC} and references
therein) it followed that 3-layer (sandwich-like) system:
insulator-metal-insulator (I/M/I) (Fig.1a) is optimal system for
realization of such mechanism. The insulating media of outer
plates of sandwich should exhibit clearly pronounced excitonic
properties. As follows from theoretical estimations
\cite{Ginz6870,PHTSC} (see, also \cite{Bard73}) for effective
interaction of conduction electrons in metallic spacer of sandwich
with excitons in its outer insulating plates, it is necessary that
metallic spacer width $d$ should be of the order of $d \sim 10
\div 30 \, \AA$.

This idea of possibility to obtain the HTSC due to exciton
mechanism with its realization, in particular, in layered
materials was undoubtely supported in a number of works on the
study of properties of HTSC cuprates in the first year after their
discovery (see, e.g., \cite{Timusk87}). However, first conclusions
on exciton mechanism of HTSC in cuprates appears to be doubted
since for measurements there were available mainly polycrystalline
samples of cuprates. Theoretical works appeared in the same year
as well as later also contained some discrepancies, up to the
total negative conclusion relative to observability of increase in
$T_c$ due to exciton mechanism in sandwich-like structures because
of relative smallness of this increase in their estimations (see,
e.g. \cite{Varma}). So, 'the question on the role of exciton
mecganism in known HTSC seems to be still quite unclear'
\cite{Ginz04,Ginz09}.\\[1mm] {\bf 3. Planar 'sandwich' in
conducting planes of HTSC cuprates ( and new iron-based
supercoductors ($T$ < $T^*$).} In spite of negative result of
search for exciton mechanism due to layered structure in cuprates
there are evidences indicating that in cuprates it is formed
another inhomogeneous I/M/I-like structure, realizing exciton
mechanism in these compounds. It is stripe structure (see, e.g.
\cite{Bia}) appearing in the CuO$_2$-plane in normal state (at $T
< T^* \approx 140 \div 200$ K, depending on compound), when the
system enters the pseudogap regime (see below). This structure
consists of the system of periodically alternating parallel
semi-insulating (spin) and conducting (charge) stripes:
I/M/I/M/I-structure (see, Fig.1b). In conducting stripes density
of mobile charge carriers is relatively high - 'metal'  while in
semi-insulating ones it is low - 'insulator'. In optimally doped
cuprates, conducting stripes ('metal') have their width near $d_M
\approx 15 \, $ \AA, and the width of semi-insulating stripes
('insulator') is near $d_I \approx 10 \,$ \AA \, \cite{Bia}. So,
this stripe structure can be considered as a series of in-plane
I/M/I/M/I-sandwiches in which it is naturally realized spatial
segregation of regions of conductivity and insulating
(semi-insulating) ones.

The feature of in-plane sandwich in this structure is that
insulating stripes-'plates' are magnetic in nature. As known,
parent (undoped) cuprate compounds are antiferromagnetic (AF)
insulators with Neel temperature $T_N \sim 300- 500$ K (depending
on compound), with ordering type of commensurate spin density wave
(SDW) - doubling of period (see, e.g. \cite{Hos09}). In doped
cuprates, the decreasing of temperature results (in contrast to
static, homogeneous AF-insulating ordering in CuO$_2$-plane in
undoped ones) in periodically modulated magnetic structure (see,
e.g. \cite{Izyu84}) arising at $T < T^* < T_N$ in this plane. This
structure, dynamical, in general case, consists of a sequence of
semi-insulating AF-domains (spin stripes) and domain walls which
are a well conducting regions (charge stripes) forming due to
bunching of excess mobile charge carriers coming to basic
CuO$-2$-plane in result of doping process. So formed AF structure
appears to be energetically more stable for given system at $T <
T^*$. The magnetic structure within these insulating domains
(stripes) are almost antiferromagnetic (length of arrows inside
these stripes in Fig.1b corresponds to the magnitude of local
magnetization), and adjacent AF insulating domains (stripes) are
in antiphase to one another \cite{KM}. Note that, according to
these calculations, the system of stripes in conducting planes of
these compounds can be realized at $T < T^*$ not only in the form
of vertical (or horizontal) sequence of alternating 'metallic' and
'insulating' stripes but in diagonal form as well as in the form
of crossing stripes when symmetry of the system is unchanged (cf.
with \cite{Fauq}), however, all effects characteristic for
HTSC-sandwich with vertical stripes (Fig.1b) are here persisted.
So, as a basis for such sublattice of quantum stripes \cite{Bia},
5-layer sandwich: I/M/I/M/I should be considered rather than
3-layer one: I/M/I/ as in Fig.1a.

The second feature of in-plane sandwich in given materials is its
dynamical character: as it was obtained already in first works on
the study of HTSC cuprates (see, e.g. \cite{KM,Bia}), stripe
structure is in general case slowly fluctuating, with
characteristic time of the order of $10^{-12}$ sec. However, since
time of interaction of conduction electrons (mobile charge
carriers) with excitons  (see (2)) is much less than
characteristic time of fluctuations of spin density then process
of this interaction can be considered as instant, i.e. it is
realized quasistatic regime what permits use, in adiabatic
approximation, for sandwich-like structures theoretical
estimations obtained before for static case (see, e.g.
\cite{Ginz76}). As for new iron-based superconductors, then
in-plane 'sandwich' in conducting planes should be formed in them
by charge stripes enriched with mobile charge carriers during
doping ('metal') alternating with spin stripes with decreased
density of mobile charge carriers (AF domains-'plates'). Though
'plates' of in-plane 'sandwich' in this case are not certainly
insulating, but rather semi-metallic ones, such possibility of
sandwich with semi-metallic, AF-ordered plates is also allowed by
Ginzburg model of sandwich \cite{Ginz6870}. \\[1mm] {\bf 4. Planar
sandwich in $\bf CuO_2$- and (FeAs-) planes as SDW/CDW system.} As
it was noted above, stripe structure in CuO$_2$-plane of HTSC
cuprates appears in the normal state when in electron energy
spectra of the system it is formed the so called pseudogap which
persists in superconducting state also ($T < T_c < T^*$), down to
lowest temperatures. Such a picture is characteristic for system
with 'interference' of dielectric and superconducting pairing in
which with decreasing temperature first at the part of the Fermi
surface it is opened a dielectric gap $\Sigma$ due to
electron-hole (e-h) pairing and only at lower temperature the
superconducting gap $\Delta$ is formed due to (e-e) pairing (see,
\cite{PHTSC,Rus}). In particular, such behaviour is characteristic
for systems of itinerant electrons with interplay of
superconductivity and magnetism \cite{Mach,Morya}. In such systems
dielectric transition occuring in the normal state at $T \le T^*$
is realized in the form of transition of the system from
spin-disordered state to the AF SDW state when at symmetric parts
of the Fermi surface it is formed SDW-dielectric gap (pseudogap)
with magnitude $\Sigma_{SDW}$ and only at lower temperature $T_c <
T^*$ it occurs a superconducting transition (it is opened a
superconducting gap $\Delta_{SC}$) so that at $T \le T_c$ two
order parameters (SDW + SC) coexist with one other in the system
\cite{Mach,M08J}. In other words, from here it follows that in
such system superconducting transition is preceded by magnetic (AF
SDW) phase transition ($T_c < T^*$). Indirect evidence for such
(AF SDW) transition in cuprates was obtained by us before, on the
basis of detailed analysis of in-plane resistive measurements in
magnetic field ($\vec H || \vec c$ ) \cite{M1}, and recently, from
detailed measurements of elastic scattering of polarized neutrons
\cite{Fauq}, it was obtained a direct evidence for such magnetic
transition in pseudogap regime for a number of HTSC cuprate
compounds.

In new iron-based superconductors, as it was already indicated
from the first measurements at polycrystals (see, e.g. \cite{Ren})
it should be observed the same picture \cite{M08P}. So,
practically just it becomes known that in undoped iron-based
compounds (semi-metal), as in cuprates, it was fixed (and have
been studied in details, in both resistive and neutron
measurements (see, \cite{Hos09,M08P}), a transition to AF SDW
state commensurate with lattice, with Neel temperature $T_N\approx
140$ K but with partial (in contrast to total one in undoped
cuprates) SDW-dielectrization of the spectra. (Here, it is
necessary to note that the problem of phase transition
metal-insulator, with dielectrization of the electron energy
spectra, was solved in well-known work \cite{KK} just for the case
of semi-metal). And indeed, in spite of more complex band
structure in iron-based compounds (as compared with the case of
cuprates) (see, e.g. \cite{Hos09}), in normal state of these
compounds, also at $T \le T^*$ it is observed pseudogap in
electron energy spectra (cf. \cite{M08P,Xu}) (which, obviously can
have more complex structure or to be a system of pseudogaps),
persisting also in superconducting state ($T < T_c < T^*$). In
addition, in literature (see, e.g. \cite{Hos09,M08P}) there are a
number of evidences indicating to SDW nature of the pseudogap
observed in new iron-based superconductors though direct
observations of quasistatic, in general case, magnetic (SDW) order
in doped compounds are absent at present (see, however, recent
work \cite{Wen}) (in cuprates, indirect (resistive) evidence for
magnetic (AF SDW) phase transition, preceding SC one, was obtained
in \cite{M1} while direct (neutron) ones only recently in
\cite{Fauq}, see above).

Since SDW-period in conducting planes of HTSC-cuprates (and new
iron-based superconductors) is (in contrast to undoped (parent)
systems) incommensurate with crystal lattice period, then
formation of SDW is accompanied by generation in these planes of
charge density wave (CDW) with one-half wavelength $\lambda_{CDW}
= \lambda_{SDW}/2$ \cite{Izyu84,KT} and thus lattice deformation
wave. As it's seen from Fig.1b, period of the SDW formed by spin
stripes (arrows), because of antiphaseness of adjacent
domain-stripes appears to be in two times larger than period of
the CDW formed by equivalent to one another charge stripes
(circles) (see, e.g., \cite{M04}). This fact indicates again (see,
Sec.3) that as in-plane sandwich in these systems it should be
taken a structure with 5 stripes I/M/I/M/I (Fig.1b) rather than
with 3 elements (I/M/I) as in the case of nonmagnetic insulating
plates (Fig.1a).
\\ [1mm]
{\bf 5. The character of excitons in HTSC-cuprates (and new
iron-based superconductors).} The exciton characteristics in AF
semi-insulating  stripes-'plates' of planar sandwich in
CuO$_2$-plane of HTSC-cuprates (see, Fig.1b) are in fact
determined by excitonic spectrum of AF-insulating CuO$_2$-plane in
parent (undoped) compounds (e.g. La$_2$CuO$_4$ and
YBa$_2$Cu$_3$O$_6$). This proposal is connected with fact that, as
it was noted above (see Sec.3), these AF semi-insulating
stripes-'plates' of planar sandwich in CuO$_2$-plane appear at $T
< T^* < T_N$ in doped compounds in fact as a consequence of
'dicretization' of initially homogeneous AF-insulating state of
CuO$_2$-plane (characteristic for undoped compounds of cuprates)
to the system of insulating antiphase spin stripes-domains
separated by charge stripes (domain walls) formed by excess mobile
charge carriers coming to CuO$_2$-plane under doping process.

As it was observed already in first measurements of optical
absorption at single crystals of cuprates (see, e.g.
\cite{BasTim05}), in dielectric region of phase ($T,x$)-diagram,
in spectra, at long-wave side of low-energy ($\varepsilon =
\hbar\omega < 3$ eV) absorption edge, in the energy range around
1.5 eV, there were observed features characteristic for excitonic
absorption. Moreover, these features appeared only at polarization
of the electic field in the incident wave in plane of
CuO$_2$-layers ($\vec E^{\sim} || ab$) which fact indicates to
planar character of excitons in cuprates (at polarization of
electric field in the incident wave normally to the plane of
CuO$_2$-layers ($\vec E^{\sim} || \vec c$ ) such features in
spectrum were absent) (the discussion on properties of same
2D-excitons and anysotropy of optical absorption in crystals see,
e.g. in \cite{AG}).

This absorption edge is provided, (see, e.g. \cite{BasTim05}), by
interband transition with charge transfer (CT-transition): O$2p
\to$ Cu $3d_{x^2 - y^2}$ in CuO$_2$-plane, with width of optical
energy gap $\Delta_{CT} \approx 1.5$ eV, so that corresponding
excitons in CuO$_2$-plane relate to the class of charge transfer
excitons (CT-excitons) (cf., \cite{Sturge,Agr03}). Such
CT-excitons are formed in CuO$_2$-plane in result of CT-transition
between two subsystems in this plane, one of which is formed by
square plaquette consisting of 4 O atoms in square vertices and Cu
atom in the centre of this square, and as another subsystem it is
considered Cu atom from adjacent plaquette to which the electron
from nearest to it O atom comes (for details, see e.g.
\cite{Wang}). In energy region to this process it corresponds
interband transition from filled $O \, 2p$-band to empty $Cu\,
3d$-band. In result, both Cu cites appear to be spinless and then
so-formed CT-excitons move across lattice in CuO$_2$-plane
coherently, without disturbance of basic AF matrix of this plane
(cf. \cite{Taoyagi85}), in contrast to single electron or hole,
movement of which in CuO$_2$-plane appears to be incoherent
leading to their localization \cite{Wang}.

Experimentally, the mobility (delocalization) of CT-excitons in
CuO$_2$-plane of these materials was established only recently,
from measurements of high resolution resonant inelastic x-ray
scattering - (RIXS)) at single crystals of parent, undoped AF
insulating compound La$_2$CuO$_4$ \cite{Coll}. Here, it is
necessary to note that possibility of using of inelastic x-ray
scattering to study the excitons of 'electronic' type, slightly
connected with lattice vibrations, was discussed in the work by
Agranovich and Ginzburg \cite{AG61} (see, also \cite{AG}), for 45
years before the work \cite{Coll}. Moreover, obtained in
\cite{AG,AG61} estimations of typical value of energy for
'electronic' exciton ($\hbar\omega \ge 1$ eV), width of
corresponding band ($\hbar \Delta \omega \ge 0.5$ eV) and even
resolution of x-ray spectrum apparatus ($\sim 0.3$ eV) in fact
coincide (!) with those for CT-excitons obtained in work
\cite{Coll}: energy $\hbar \omega \le 1.5$ eV, band width $\sim
0.5$ eV and 'high resolution of apparatus' (synchrotron) $\sim
300$ meV.

This study demonstrates that CT-excitons in CuO$_2$-plane are
mobile quasiparticles with quadratic dispersion (in contrast to
the case of localized excitons in other isostructural compounds
(e.g. La$_2$NiO$_4$)). In addition, the whole mass of CT-exciton
$M = m_e^* + m_h^*$ , determining its motion as a whole is equal,
according to estimation in Wannier-Mott exciton model, $M = 3.5 \,
m_e$, that also is in agreement with estimations used before
\cite{AG}. Here, $m_e$ is the mass of the free electron, $m_e^*$
and $m_h^*$ are, respectively, effective masses of quasiparticle
and quasihole forming an exciton.\\ [1mm] {\bf 6. The effective
parameters of a BCS model for planar CT-excitons in HTSC-cuprates
(and new iron-based superconductors).} So, as follows from these
experiments, in material of insulating 'plates'-stripes of planar
sandwich in CuO$_2$-plane which, in fact, are result of
'discretization' of uniformly AF-ordered insulating CuO$_2$-plane
(characteristic for undoped compounds) (see, above, Sec.3) can
propagate CT-excitons with characteristic energy $\hbar \omega \le
1.5$ eV, providing the possibility for superconducting pairing of
conduction electrons (mobile charge carriers) in 'metallic' spacer
of sandwich at higher $T_c$ due to exchange by these CT-excitons.
The characteristic temperature $\Theta_{ex}$, corresponding to
this energy, is of the order of $\sim 10000-15000$ K, that is in
agreement with estimations for this mechanism, see (2). Such
agreement permits estimate effective parameter $g$, determining
the measure of attraction of conduction electrons (mobile charge
carriers) in 'metallic' spacer. From expression (1) for critical
temperature $T_c$ it follows that $g \sim 1/ln(\Theta_{ex}/T_c)$ ,
and, for example, for maximum critical temperature $T_c^{max}
\approx 90$ K in optimally doped YBa$_2$Cu$_3$O$_{7-\delta}$ we
have: $g \sim 0.2$. Analogously, for La$_{2 - x}$Sr$_x$CuO$_4$
($T_c^{max} \approx 40$ K), $g \sim 0.18$.

Further, in undoped iron-based compounds, as follows from recent
measurements of optical absorbtion at single crystals (see, e.g.
\cite{Bor08,Qazi}), features (under conditions of finite
conductivity of these compounds) characteristic for excitonic
absorption in lightly doped cuprates were observed in spectrum at
energy $\hbar \omega \sim 0.65$ eV, and, as in cuprates, at
long-wave side of absorption edge, which can correspond (in
analogy with cuprates) to CT-transition (cf. with \cite{Zaan}) in
conducting planes of these compounds. This energy corresponds to
characteristic temperature of excitons of the order of
$\Theta_{ex} \sim 7000$ K, what, at maximum reached $T_c^{max}
\approx 55$ K, leads to the value of interaction constant also of
the order $g \sim 0.2$.

Since, in real HTSC single crystals, even at exciton mechanism, it
is necessary to take into account the interaction of conduction
electrons (mobile charge carriers) with phonons (cf.
\cite{Ginz00}), then obtained above magnitudes of effective
parameter $g \sim 0.2$, which are in agreement on the order of
magnitude with calculated magnitudes for electron-phonon
interaction in cuprates (see, e.g., \cite{Heid}), and in new
iron-based superconductors (see, e.g., \cite{Dolg}), indicate to
moderate force of both electron-phonon and electron-exciton
interactions in both classes of these compounds. Thus, correlation
of magnitudes of effective parameters $\Theta$ and $g$ in BCS
model (1) obtained for both classes of compounds (see, also (2))
denotes in fact that high values of critical temperature of
superconducting transition in these compounds are provided by
combination of both parameters in (1): sharp rise of
pre-exponential factor $\Theta$ due to mobile, planar CT-excitons
and increase of density of states in exponent index $g$, due to
partial SDW-dielecrization of electron energy spectra (accompanied
by formation of in-plane sandwich in conducting planes) in the
normal state.

It is interesting that in fact to the same conclusion that high
critical temperature of superconducting transition in cuprates is
a result of combination of contributions from excitons and
moderate phonon mechanism came and author of pioneer work
\cite{Little64} Little with coauthors \cite{Little07}. In this
recent work, with using of precisious measurements of optical
absorption in thin crystalline films of Tl$_2$Ba$_2$CaCu$_2$O$_8$
it was studied frequency dependence of relation of magnitudes of
optical conductivity in superconducting $\sigma_s$ and normal
$\sigma_n(T \ge Tc)$) states:
$Re[\sigma_s(\omega)/\sigma_n(\omega)]$ . At experimental curve
there was observed a declination from known calculated dependence
in frames of the BCS theory \cite{MBard58}, in particular, at
energy $\hbar \omega \approx 1.2$ and $1.7$ eV, which with taking
into account measured RIXS-spectra (cf. with above
\cite{AG,Coll,AG61}) were treated as exciton peaks, corresponding
to electronic excitations in the copper subsystem (cf., however
\cite{BasTim05}). The characteristic temperatures corresponding to
obtained energies are in agreement on the order of magnitude
($\Theta_{ex} \sim 10^4$ K), with both obtained above
corresponding temperatures in CuO$_2$-planes of LSCO and YBCO and
with (2). And value of $T_c \approx 105 K$ for their films results
also in $g \sim 0.2$.
\\[1mm]
{\bf 7. The onset temperature of superconducting transition
($T_c^{onset}$) at exciton mechanism in cuprates (and new
iron-based superconductors.} Though, principal scheme of
manifestation of the Little-Ginzburg exciton mechanism in doped
compounds of cuprates (and new iron-based superconductors) are
already above described, however, in real systems there is exists
a number of features caused by additional factors determining
behaviour of the system of mobile charge carriers in the normal
state and decreasing thus real temperature of superconducting
transition in the system. So, for example, it is necessary to take
into account role of fluctuations of magnetic (SDW) order
parameter (amplitude spin fluctuations of local spin density)
characteristic for magnetic systems of itinerant electrons (see,
e.g. \cite{Morya}) which gives rise, in particular, to additional
(to phonon one $\rho_{ph}(T))$ contribution $\rho_m(T)$ to the
total resistivity $\rho_{tot}(T)$ ( $\rho_{tot}(T)$ =
$\rho_{ph}(T))$ + $\rho_m(T)$, usual approach for magnetic metals,
see, e.g. \cite{M04}). This contribution $\rho_m(T)$  to
resistivity disappears (characteristic time of fluctuations of
spin subsystem is noticeably larger than characteristic time of
electronic processes) at temperatures determined by parameters of
real system.

Experimentally, onset temperature of superconducting transition
$T_c^{onset}$ for concrete HTSC single crystal can be determined
from point of intersection of temperature dependence of in-plane
resistivity $\rho(T)$ with well known universal Gruneisen-Bloch
curve determining temperature dependence of phonon contribution in
resistivity $\rho_{ph}(T)$ in most metals \cite{Zai} (for more
details, see \cite{M1,M08P}). At this temperature, the behaviour
of the system of mobile charge carriers in 'metallic' spacer (M)
of in-plane sandwich begins to be determined (without taking into
account the influence of its 'insulating' plates (I)) only by
interaction with phonons, which, however, as follows from
expression (1), appears to be not so strong for Cooper pairing of
mobile charge carriers at given temperature, even under conditions
of increased density of states at the ends of SDW-dielectric gap
(pseudogap) at the Fermi surface ($T < T^*$) \cite{Rus,M08J}.
Because of this, 'insulating' plates (I) of in-plane sandwich play
here a crucial role - it is turned on the interaction of mobile
charge carriers in 'metallic' spacer (M) of in-plane sandwich with
CT-excitons in its 'insulating' plates (I) (with essentially wider
region of energy for interaction near the Fermi surface than only
at phonon mechanism) leading to their effective Cooper pairing and
transition of the system to superconducting state at given
(unreally high for only phonon mechanism) temperature.

Then, as follows from above, excitons ('polarization waves'
\cite{Ginz64,Ginz6870,PHTSC,AG}) from insulating stripe-'plate' of
in-plane sandwich (Fig.1b) can interact with mobile charge
carriers in its 'metallic' stripe-spacer, even from two sides,
providing thus additional mutual interaction of mobile charge
carriers due to exchange by mobile, planar CT-excitons giving rise
to formation of Cooper pairs and superconducting transition at
temperature significantly higher than phonon limit ( $T_c^{ph} \le
40$ K, see, e.g., \cite{PHTSC}). On the other hand, it is
necessary also note that in the systems under consideration, since
width of 'insulating' plates (I) of in-plane sandwich satisfies
the conditions for Josephson tunneling, mobile charge carriers
from adjacent 'metallic' spacers (M) are able to form a Cooper
pairs (due to exchange by mobile, planar CT-excitons from
'insulating' plates (I) between them) what is consistent with the
fact that e.g. in YBCO-system the distance between 'metallic'
spacers (M) in planar sandwich - CDW wavelength ($\lambda_{CDW}$),
is equal to coherence length in CuO$_2$-plane ($\xi_{ab}$):
$\xi_{ab} \approx \lambda_{CDW} \approx 25 \, \AA$ \cite{M1,M04}.
So, adjacent AF ordered 'insulating' plates-domains (I) of
in-plane sandwich, because of their antiphasness can provide
additional stabilization of the system (at given (high)
temperature) due to co-phasing of spins of mobile charge carriers
in its adjacent equivalent 'metallic' spacers (M) (domain walls).
\begin{figure}
\includegraphics[width=17.0 cm]{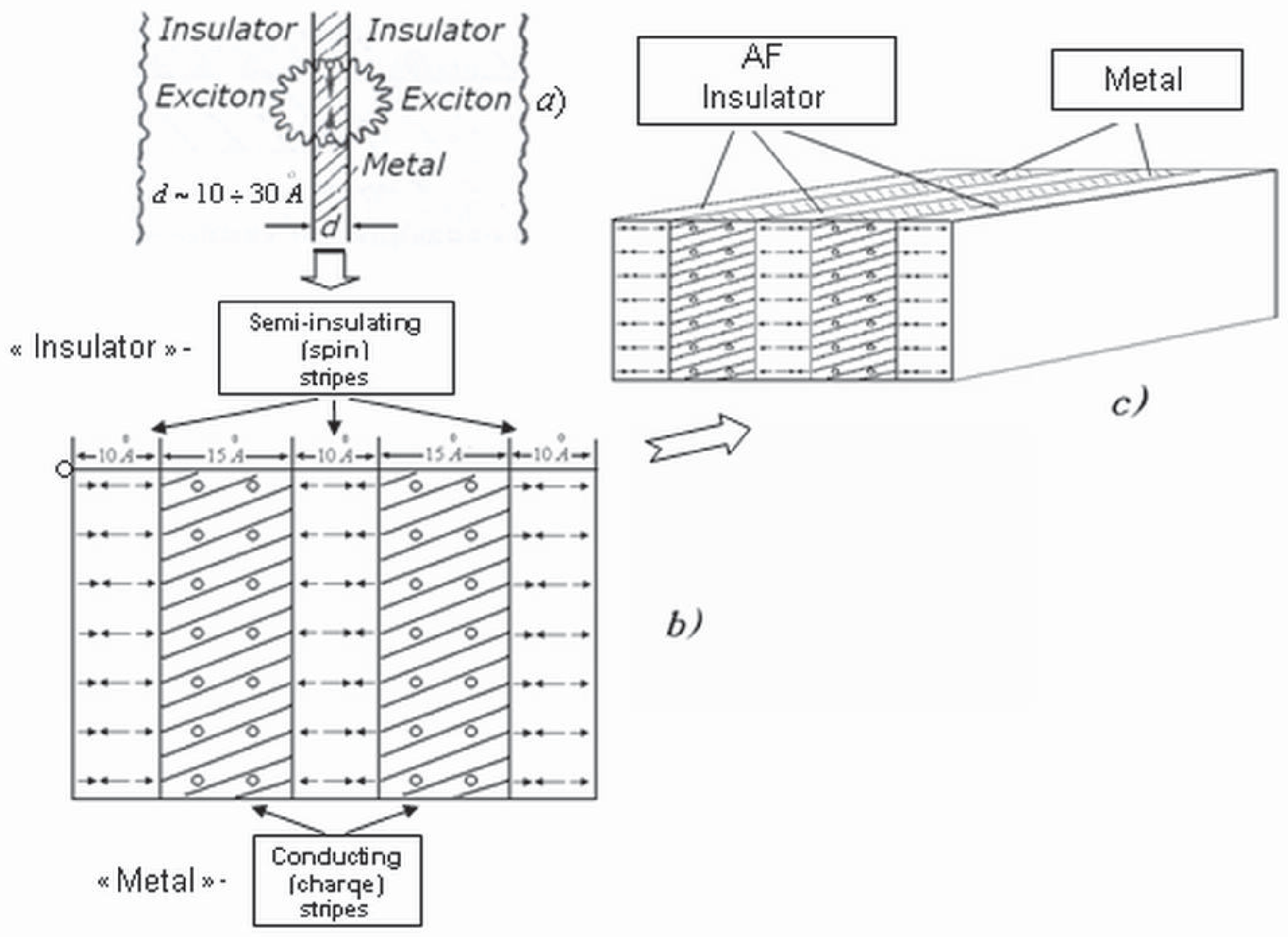}
\caption{The sketch of HTSC-sandwich-like system. a) 3-layer
Ginzburg HTSC-sandwich with insulating outer plates and metallic
spacer (shaded) between them; d is the thickness of the metallic
spacer; two circles in the middle of metallic spacer correspond to
conduction electrons which are attracted to one other via exchange
by virtual excitons (wavy lines), see, e.g. \cite{Ginz6870}. b)
5-layer planar 'HTSC-sandwich' (stripe structure) in the CuO2
plane of cuprates: insulating  'plates' (semi-insulating, spin
stripes) are AF-ordered (arrows indicate direction of local
magnetization inside the 'plate', the length of arrow corresponds
to magnitude of this magnetization, after \cite{KM});
'metallic-spacer' (conducting, charge stripes) regions are shaded,
circles in these regions correspond to mobile charge carriers. The
'insulating-plate' and 'metallic-spacer' widths correspond to the
cuprates with optimal doping \cite{Bia}. Note that nearest
insulating 'plates' are in antiphase to one other, and the spin
density wave (SDW) formed by them as well as accompanying it the
charge density wave (CDW) formed by equivalent 'metallic spacers'
are incommensurate with the lattice period (see, e.g. \cite{M0006}
and references therein); c) artificial multilayered structure
metal/AF-insulator, cross section of which corresponds to in-plane
sandwich in Fig.1b.}\label{disp2}
\end{figure}

Of course, presented here picture is schematic enough and needs in
further development however, it, at least, permits describe from
unified viewpoint so-called 'anomalous' behavior of these
superconductors and obtain self-consistent, quantitative
estimations of critical parameters \cite{M1,M04}. It also
indicates possible ways of increasing of critical temperature
(RTSC) in similar, layered AF-ordered compounds with CT-transition
in plane of layers as well as in conducting, artificial
inhomogeneous sandwich-like systems, where insulating material of
plates is AF-ordered \cite{M0006}, what permits to hope for
further increase of critical temperature in such compounds in near
future. \\[1mm] {\bf 8. Possibility of creation of new artificial
HTSC (RTSC) superlattices with AF-insulating layers.} The picture
of stripes in CuO$_2$-planes of HTSC and determination of their
widths \cite{Bia} led in beginning of 90th to attempts to
reconstruct such stripe structure with critical temperature $T_c
\sim 300$ K (RTSC) artificially, on the basis of metallic
structure at the atomic limit (see \cite{BiaPat}), which can be
realized with using of molecular beam epitaxy, evaporation,
lithography, chemical sintesis, electrochemical deposition etc.
Such superconducting heterostructure is formed by superlattice of
quantum elements (wires by thikness $ \sim 10 \, \AA$) of one
material (metal) incorporated between another ones (insulator),
which provides periodic potential barrier for electrons in first
one.

However, in this picture the insulating elements of such
superlattice are considered only as creating periodic potential
for metallic elements without taking into account their magnetic
nature. Of course, from viewpoint of exciton mechanism such
structure (in 2D version) can be considered as planar 'sandwich'
but only as 3-layer I/M/I heterostructure (see Fig.1a). The taking
into account the magnetic nature of 'insulating' stripes permits
propose a number of new ways to model HTSC cuprates. Most direct
of them is creation of artificial magnetic (insulating) stripes in
planes parallel to conducting planes of HTSC or (and)
AF-insulating layers with corresponding metallic ones realizing
SDW/CDW structure with charge and AF-insulating stripes in
cross-section, see Fig.1c.

The works of such type on HTSC, at present time, as known, are
performed. First, it is necessary to note experiments with
atomically plane thin HTSC films and heterostructures
(multilayered structures, superlattices) \cite{Bozh08}. For these
experiments, with using of molecular beam epitaxy there were
created cuprate HTSC-heterostructures with atomically smooth
surfaces and interfaces. Further, in the process of growth of
structure, in real-time regime, with using of special precisious
apparatus it was controlled the crystal structure of the surface,
chemical composition of surface layers etc. what permitted
directly correct a technological regime. There were studied the
effects of contact of thin layers of HTSC cuprates with different
level of doping (undoped, underdoped, optimally doped, overdoped).
In particular, it was obtained superconductivity with high $T_c$
at the boundary of double layers consisting of two
non-superconducting (overdoped and underdoped) cuprates (LSCO).

To the same class of experiments, obviously, can be related also
that in \cite{Locq}, where film of LSCO, with width $ \sim 150 \,
\AA$  was grown with using of molecular epitaxy at SrLaAlO$_4$
substrate which lattice period was slightly different from lattice
period of the film grown. Incommensurateness of lattices led to
twofold increase of $T_c$ as compared with bulk sample. Increase
of $T_c$ in these cases can be caused by secondary effect (see
above, Sec.4) namely, creation of lattice deformation wave in the
sample of HTSC cuprate at given temperature which can lead to
generation of CDW and hence SDW in conducting planes (formation of
stripe structure (a series of in-plane sandwiches)) and thus to
both increase of density of states at the Fermi surface and
appearance of conditions for effective interaction of mobile
charge carriers in 'metallic' stripes and mobile CT-excitons in
'insulating' stripes, and, hence, to superconducting transition at
higher temperature due to combination of these two factors.

In general case, for creation of artificial sandwich-like
structures with AF-insulating layers-plates (see Fig.1c and
\cite{M0006}) it can be used a technology of layer-by-layer
evaporation of alternating layers metal/insulator: such methods
with evaporation of layers with width  $\sim 10 \, \AA$ are used
at production of x-ray mirrors (see, e.g. \cite{IFM}), structures
with giant magnetoresistance (see, e.g. \cite{Buss99}), and namely
superconducting heterostructures metal/semiconductor \cite{Fog96}.
\\[1mm] {\bf 9. Conclusion.}  So, performed analysis demonstrates
that in both cuprates and new iron-based superconductors there are
satisfied all conditions of Ginzburg, optimal for realization of
exciton mechanism in HTSC-sandwich (Fig.1) which were formulated
by him in the end of 60th \cite{Ginz6870}: \\[1mm]
 \quad $d \sim 10-30 \, \AA, \quad n \sim
10^{18}-10^{23}$ cm$^{-3}$, \quad $\hbar \omega_{ex} \sim 0.3-3$
eV, \quad   $\hbar \Delta \omega \sim 0.1 - 3$ eV.  \\

Moreover, appearance of high-temperature (middle-temperature)
superconductivity already in second class of layered AF systems
with transition of parent (undoped) system to commensurate SDW
state with total (cuprates) or partial (iron-based compounds)
SDW-dielectrization of electron energy spectrum and appearance in
the system of mobile, planar CT-excitons is important in search
for HTSC and RTSC in such systems which can be at present
considered as most promising. The key moment in such search can be
study of layered systems of transition metals with high value of
energy of CT-transition in plane of layers which in parent
(undoped) state are instable relative to transition to the state
with commensurate CDW when in the system it occurs a total
(CT-insulator) or partial (CT-semi-metal) SDW-dielectrization of
electron energy spectra. The search for such compounds obviously
should be performed with using of measurements of optical
absorption at single crystals, at polarization of electric vector
in incident wave parallel to layers ($\vec E^{\sim}|| ab$). After
selection of promising compounds it can be realized their doping
to perform the test for superconductivity with using of
measurements of susceptibility or resistivity at $T \le 300$ K.
Since materials of such type are well known enough and after
discovery of new iron-based superconductors, the investigation on
search for similar compounds is coming intensively, then, in
principle, it is possible to wait an appearence of RTSC in them in
near future. \\

The author is thankful to V.L.Ginzburg for a critical reading of
the work.


\begin{thebibliography}{99}
\bibitem{Hos08} Kamihara Y. et al. {\it J.Am.Chem.Soc.} {\bf 130} 3296 (2008)

\bibitem{Hos09} Ishida K. et al. {\it J.Phys.Soc.Jpn.} {\bf 78} 062001 (2009).

\bibitem{Ginz09} Ginzburg V.L. {\it On Superconductivity and Superfluidity}
(Springer-Verlag, Berlin, 2009).

\bibitem{Ginz04} Ginzburg V.L. {\it Usp.Fiz.Nauk} {\bf 174} 1240 (2004)

\bibitem{Little64} Little W A {\it Phys.Rev.} {\bf 134} A1416 (1964)

\bibitem{Ginz64} Ginzburg V.L. {\it Zh.Exp.Teor.Fiz.} {\bf 47} 2318 (1964); {\it Phys.Lett.}
{\bf 13} 101 (1964)

\bibitem{Ginz6870} Ginzburg V.L. {\it Usp.Fiz.Nauk} {\bf 95} 91 (1968); {\bf 101} 185 (1970)

\bibitem{Ginz76} Ginzburg V.L. {\it Usp.Fiz,Nauk} {\bf 118} 315 (1976)

\bibitem {PHTSC} {\it Problem of High Temperature Superconductivity} (Eds. V.L.Ginzburg
and D.A.Kirzhnitz, Nauka Press, Moscow, 1977) [New York:
Consultants Bureau, 1982]

\bibitem{Bard73} Allender D., Bray J., Bardeen J.   {\it Phys.Rev.} B {\bf 7} 1020  (1973)

\bibitem{Timusk87} Kamaras K. et al. {\it Phys. Rev. Lett.} {\bf 59} 919 (1987)

\bibitem{Varma} Varma C.M., Schmitt-Rink S., Abrahams E.  {\it Sol.St.Commun.} {\bf 62}
681 (1987)

\bibitem{Bia} Bianconi A. et al. {\it Phys.Rev.Lett} {\bf 76} 3412 (1996)

\bibitem{Izyu84} Izyumov Yu.A. {\it Usp.Fiz.Nauk} {\bf 144} 439 (1984)

\bibitem{KM} Kato M. et al.  {\it J.Phys.Soc.Jap.} {\bf 59} 1047 (1990)

\bibitem{Rus} Rusinov A.I. et al.  {\it Zh.Exp.Teor.Fiz.} {\bf 65} 1984 (1973)

\bibitem{M08P} Mazov L.S.  arXiv:0805.4097 (2008)

\bibitem{Fauq} Fauque B. et al. {\it Phys.Rev.Lett.} {\bf 96}, 197001 (2006)

\bibitem{Mach} Machida K.  {\it Appl.Phys.} A {\bf 35} 193 (1984)

\bibitem{Morya} Moriya E. {\it Spin fluctuations in magnetics with itinerant electrons}
(Springer-Verlag, Berlin, 1985).

\bibitem{M08J} Mazov L.S.   {\it J.Supercond.Nov.Magn.} {\bf 20} 758  (2008);
{\it J.Supercond.} {\bf 18} 713 (2005).

\bibitem{M1} Mazov L.S. {\it Low Temp.Phys.}  {\bf 17} 1358 (1991); see, also in {\it Progress in High
Temperature Superconductivity}, vol.32, (Eds.: A.G.Aronov, A.I.
Larkin and V.S. Lutovinov, World Scientific, Singapore) (1992),
p.605-610;

\bibitem{Ren} Zhi-An Ren et al.  {\it arXiv:0803.4243} (2008)

\bibitem{KK} Keldysh L.V., Kopaev Yu.V. {\it Sov.Phys.-Solid State} {\bf 6} 2792 (1964)

\bibitem{Xu} Xu Y-M  et al. arXiv:0905.4467 (2009)

\bibitem{Chen} Chen, H. et al.  {\it Europhys.Lett.} {\bf 85}, 17006 (2009).

\bibitem{KT} Kulikov N.I., Tugushev V.V.  {\it Usp.Fiz.Nauk} {\bf 144} 643 (1984)

\bibitem{M04} Mazov L S  {\it Phys.Rev.} B {\bf 70} 054501 (2004); see, also in
{\it Superconductivity Research at the Leading Edge} (Ed. P.S.
Lewis) (New York: Nova Science Publishers, Inc., 2004)  p.1-23

\bibitem{BasTim05} Basov D N, Timusk T  {\it Rev.Mod.Phys.} {\bf 77} 721 (2005)

\bibitem{AG} Agranovich V.M., Ginzburg V.L. {\it Crystal Optics
with Taking into Account of Spatial Dispersion and the Theory of
Excitons} (Moscow: Nauka Press, 1965); [Interscience Publishers,
New York, 1966]

\bibitem{Sturge} Sturge M.D. in: {\it Excitons} (Eds.: E.I.Rashba and M.D.Sturge)
(Moscow: Nauka Press, 1985), p.9 (in Russian)

\bibitem{Agr03} Knoester I., Agranovich V.M.  {\it Thin films and nanostructures}
{\bf 32}, 1 (2003)

\bibitem{Wang} Wang Y.Y. et al. {\it Phys.Rev.Lett.} {\bf 77} 1809 (1996)

\bibitem{Taoyagi85} Tanabe Yu., Aoyagi K. in {\it Excitons} {Eds.:
E.I.Rashba and M.D.Sturge} (Moscow, Nauka Press, 1985), p.424 (in
Russian)

\bibitem{Coll} Collart E. et al. {\it Phys.Rev.Lett.} {\bf 96} 157004 (2006)

\bibitem{AG61} Agranovich V.M., Ginzburg V.L. {\it Zh.Exp.Teor.Fiz.} {\bf 40} 913 (1961)

\bibitem{Bor08} Boris A.V. et al. arXiv:0806.1732 (2008)

\bibitem{Qazi} Qazilbash M.M. et al. arXiv:0808.3748 (2008)

\bibitem{Zaan} Zaanen J., Sawatzky G.A., Allen J.W.  {\it Phys.Rev.Lett.}
{\bf 55} 418 (1985)

\bibitem{Ginz00} Ginzburg V.L. {\it Usp.Fiz.Nauk} {\bf 170} 619 (2000)

\bibitem{Heid} Heid R. et al. {\it Phys.Rev.Lett.} {\bf 100}, 137001 (2008)

\bibitem{Dolg} Boeri L., Dolgov O.V., Golubov A.A.  {\it Phys.Rev.Lett.} {\bf 101}
 026403 (2008)

\bibitem{Little07} Little W.A. et al. {\it Physica C Superconductivity}
{\bf 460-462}  40  (2007)

\bibitem{MBard58} Mattis D.C., Bardeen J.  {\it Phys.Rev.}  {\bf 111} 412 (1958)

\bibitem{Zai} Ziman J.M.  {\it Electron and phonons} (Oxford at the Clarendon Press, 1960)

\bibitem{M0006} Mazov L.S.  {\it Int.J.Mod.Phys.} B {\bf 14}, 3577
(2000);see, also in  {\it Symmetry and Heterogeneity in High
Temperature Superconductors} V.214 (Ed. A.Bianconi) (Netherlands
(Dordrecht): Springer (NATO Science Ser. II, 2006) p.217

\bibitem{Wen} Wen J. et al. arXiv:0906.3774 (2009)

\bibitem{BiaPat} Bianconi A. {\it European Patent} N.0733271 “High Tc Superconductors
Made by Metal Heterostructures at the Atomic Limit” toward Room
Temperature Superconductivity: (priority date 7 Dec 1993),
published in European Patent Bulletin 98/22 (1998)

\bibitem{Bozh08} Bozovich I.  {\it Usp.Fiz.Nauk} {\bf 178}  179  (2008)

\bibitem{Locq} Locquet  J.-P. et al.  {\it Nature} {\bf 394}  453  (1998)

\bibitem{IFM} Andreev S.S.  et al.  {\it SPIE Proc.} {\bf 3406} 45 (1998)

\bibitem{Buss99} Bussmann K. et al. {\it Appl.Phys.Lett.} {\bf 75} 2476 (1999)

\bibitem{Fog96} Fogel N.Ya. et al.  {\it Low Temp.Phys.}  {\bf 22}  359
(1996) \\ [1mm]
\end{thebibliography}
\end{document}